\documentclass[superscriptaddress,aps,prl,twocolumn,showpacs,preprintnumbers,amsmath,amssymb,floatfix,groupedaddress]{revtex4}

\usepackage{graphicx}
\usepackage{dcolumn}
\usepackage{bm}

\begin{document}

\title{Coherent Quantum Optical Control with Subwavelength Resolution}

\author{Alexey V. Gorshkov}
\author{Liang Jiang}
\author{Markus Greiner}
\affiliation{Physics Department, Harvard University, Cambridge, Massachusetts 02138, USA}
\author{Peter Zoller}
\affiliation{Institute for Quantum Optics and Quantum Information of the Austrian Academy of Sciences, A-6020 Innsbruck, Austria}
\author{Mikhail D. Lukin}
\affiliation{Physics Department, Harvard University, Cambridge, Massachusetts 02138, USA}

\date{\today}

\begin{abstract}
We suggest a new method for quantum optical control with nanoscale resolution. Our method allows for coherent far-field manipulation of individual quantum systems with spatial selectivity that is not limited by the wavelength of radiation and can, in principle, approach a few nanometers. The selectivity is enabled by the nonlinear atomic response, under the conditions of Electromagnetically Induced Transparency, to a control beam with intensity vanishing at a certain location. Practical performance of this technique and its potential applications to quantum information science with cold atoms, ions, and solid-state qubits are discussed.
\end{abstract} 


\maketitle

Coherent optical fields provide a powerful tool for coherent manipulation of a wide variety of quantum systems. Examples range from optical pumping, cooling, and quantum control of isolated atoms \cite{bloch05jaksch04,science02} and ions \cite{wineland98} to manipulation of individual electronic and nuclear spins in solid state \cite{wrachtrup06,imamoglu06}. However, diffraction sets a fraction of the optical wavelength $\lambda$ as the fundamental limit to the size of the focal spot of light \cite{bornwolf}. This prohibits high-fidelity addressing of individual identical atoms if they are separated by a distance of order $\lambda$ or less. In this Letter, we propose a method for coherent optical far-field manipulation of quantum systems with resolution that is not limited by the wavelength of radiation and can, in principle, approach a few nanometers.

\begin{figure}[b]
\includegraphics[scale = 0.37]{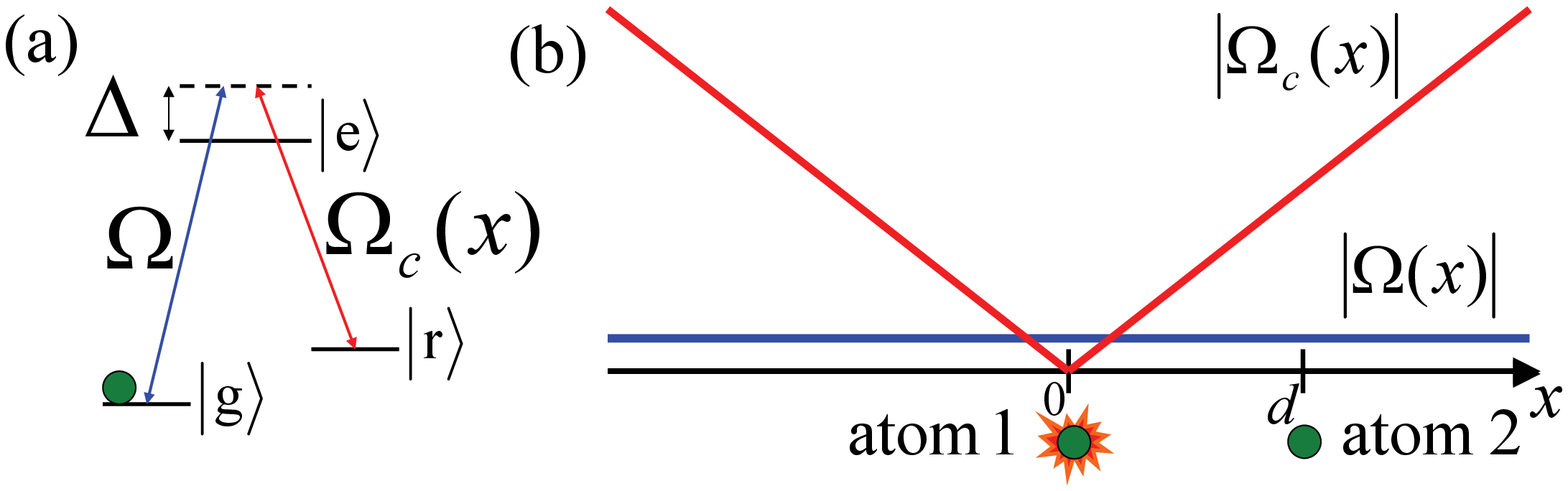}
\caption{(color online) (a) 3-level atom prepared in state $|g\rangle$ and coupled at two-photon resonance to a spatially uniform probe field with Rabi frequency $\Omega$ and a spatially varying control field with Rabi frequency $\Omega_c(x)$. (b) Schematic of the setup: atom 1, at a node of the control field, responds to the probe, while atom 2, a distance $d$ away, is subject to a large control field $\Omega_c(d) \gg \Omega$ and does not respond to the probe. \label{schematic}}
\end{figure}

Our method for coherent sub-wavelength manipulation is based on the nonlinear atomic response produced by  so-called dark resonances \cite{scully97}. The main idea can be understood using the three-state model atom shown in Fig.\ \ref{schematic}(a). Consider two such atoms, atom 1 and atom 2, positioned along the $x$-axis at $x = 0$ and $x = d$, respectively, as shown in Fig.\ \ref{schematic}(b). Assume that they are prepared in the ground state $|g\rangle$ and then illuminated by the probe field with wavelength $\lambda$ and Rabi frequency $\Omega$. For $d \ll \lambda$, one cannot focus the probe on atom 1 without affecting atom 2 and other  neighboring atoms. Let us suppose that $\Omega$ is uniform over the distance $d$. In addition, prior to turning on the probe, we turn on a  two-photon-resonant spatially varying control field (e.g.\ a standing wave) of wavelength $\lambda' = 2 \pi/k'$ that vanishes at $x = 0$ (i.e.\ has a node) and has Rabi frequency $\Omega_c(x) \approx \Omega_0 k' x$ for $k' x \ll 1$. Destructive interference of excitation pathways from $|g\rangle$ and $|r\rangle$ up to $|e\rangle$ ensures that the so-called dark state $|\textrm{dark}(x)\rangle = (\Omega_c(x) |g\rangle - \Omega |r\rangle)/\sqrt{\Omega^2_c(x) + \Omega^2}$ is decoupled from both optical fields \cite{scully97}. It is the sharp nonlinear dependence of $|\textrm{dark}(x)\rangle$ on $\Omega_c(x)$ that allows for sub-wavelength addressability. In particular, for atom 1 at $x = 0$, $|\textrm{dark}(x)\rangle = -|r\rangle$, so that atom 1 prepared in state $|g\rangle$ responds to the probe light in the usual way. On the other hand, for all $x$ such that $\Omega_c(x) \gg \Omega$, $|\textrm{dark}(x)\rangle \approx |g\rangle$. The space interval around $x = 0$, in which the ground state $|g\rangle$ is not dark, therefore, has width $\sim \Omega/(\Omega_0 k')$ and can thus be made arbitrarily small by increasing the overall intensity of the control ($\propto \Omega_0^2$). In particular, atom 2 at $x = d$ prepared in $|g\rangle$ will not respond to the probe provided $\Omega_0 \gg \Omega/(k' d)$.

This selective sub-wavelength addressability can be used in a variety of ways. For example, one can accomplish  selective  state manipulation  of proximally spaced qubits via spatially selective stimulated Raman transitions. In combination with dipole-dipole interactions, our technique can be used, for $d \ll \lambda$, to generate an efficient two-qubit gate between pairs of atoms. One can implement selective fluorescence detection \cite{wineland98} of the internal state of an atom if $|g\rangle-|e\rangle$ corresponds to a cycling transition (this is possible either if $|r\rangle$ is above $|e\rangle$ or if spontaneous emission from $|e\rangle$ into $|r\rangle$ is much slower than into $|g\rangle$). Finally, one can perform spatially selective optical pumping of individual atoms. Addressability with $d \ll \lambda$ will be important for arrays of quantum dots \cite{imamoglu06} or optically active defects \cite{wrachtrup06} in solid state, where $d \ll \lambda$ is often needed to achieve coupling \cite{meijer06lukin00}. Moreover, our technique enables highly desirable high-fidelity micron-scale manipulation of atoms in optical lattices with $d = \lambda/2$ \cite{bloch05jaksch04} and ions in linear Paul traps with $d < 5$ $\mu$m \cite{wineland98} (for ions, small $d$ is desirable as it accompanies large vibrational frequencies \cite{wineland98}). Below, we analyze in detail selective coherent state manipulation and then estimate manipulation errors using realistic experimental parameters.  

Before proceeding, we note important prior work. Our approach is an extension of incoherent nonlinear techniques used in atom lithography \cite{prentiss98} and biological imaging \cite{hell07}. The nonlinear saturation of EIT response that forms the basis of the present work has already been used for the realization of stationary pulses of light \cite{bajcsy03} and has been suggested for achieving subwavelength localization of an atom in a standing wave (\cite{sahrai05, agarwal06} and references therein). Finally, alternative approaches to solving the addressiblity problem exist that use Bessel probe beams with nodes on all but one atom \cite{saffman04}, place atoms into traps separated by more than $\lambda$ \cite{largeseparation}, and resolve closely spaced atoms spectroscopically \cite{thomas89} by applying spatially varying magnetic fields \cite{schrader04} or light shifts \cite{thomas93zhang06lee07}.

\begin{figure}[b]
\includegraphics[scale = 0.39]{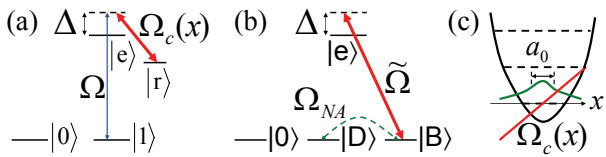} 
\caption{(color online) Single-qubit phase gate on atom 1. (a) Atom 1 ($\Omega_c(0)=0$) or atom 2 ($\Omega_c(d) \neq 0$). (b) Atom 2 using basis states $\left\{|D\rangle,|B\rangle\right\}$ in place of $\left\{|1\rangle,|r\rangle\right\}$. (c) Schematic of imperfect localization of atom 1: parabolic trapping potential $m w^2 x^2/2$ with three lowest energy levels indicated, ground state wavepacket of width $a_0$, and control field $\Omega_c(x) \approx \Omega_0 k' x$. \label{rotations}}
\end{figure}

As a specific example, we now analyze in detail a spatially selective single-qubit phase gate, $|0\rangle \rightarrow |0\rangle$, $|1\rangle \rightarrow e^{i \phi} |1\rangle$,  on a qubit encoded in stable atomic states $|0\rangle$ and $|1\rangle$ of one atom in the presence of a proximal neighbor (Fig.\ \ref{rotations}). Consider atoms 1 and 2 that have a tripod configuration shown in Fig.\ \ref{rotations}(a). We assume that the 
optical transitions from the metastable states $|0\rangle$, $|1\rangle$, and $|r\rangle$ up to $|e\rangle$ are separately addressable via polarization or frequency selectivity. By turning on a probe field with Rabi frequency $\sim \Omega$, wavelength $\lambda = 2 \pi/k$, and detuning $\Delta \gg \Omega$ for a time $\tau \propto \Delta/\Omega^2$, we would like to apply a $\pi$-phase on state $|1\rangle$ of qubit 1 via the ac Stark effect. To minimize errors discussed below, we turn  $\Omega$ on and off not abruptly but adiabatically (e.g.\ a linear ramp up from zero immediately followed by a linear ramp down to zero). To shut off the phase shift on the nearby qubit 2, we apply, at two-photon resonance with $\Omega$, a spatially varying control field with Rabi frequency $\Omega_c(x) \approx \Omega_0 k' x$ for $k' x \ll 1$. We assume the control is turned on before and turned off after the probe pulse. For the moment we treat atoms as point particles and return to the question of finite extent of the atomic wave packets below. 

The gate error on atom 1 due to spontaneous emission can be estimated as $\tau \gamma \rho_e \sim \tau \gamma (\Omega/\Delta)^2 \sim \gamma/\Delta$, where $\rho_i$ is the population of state $|i\rangle$ and where we assume for simplicity that all transitions are radiatively broadened and that the decay rate of $|e\rangle$ is $2 \gamma$. To investigate the effect on atom 2, we define dark and bright states for this atom as $|D\rangle = (\Omega_c |1\rangle - \Omega |r\rangle)/{\tilde \Omega}$ and $|B\rangle = (\Omega |1\rangle + \Omega_c |r\rangle)/{\tilde \Omega}$, where $\tilde \Omega = \sqrt{\Omega_c^2 + \Omega^2}$ and $\Omega_c = \Omega_c(x=d)$ (see Fig.\ \ref{rotations}(b)). Since $|D\rangle = |1\rangle$ at the beginning and at the end of the probe pulse (i.e.\ when $\Omega = 0$), the phase gate will be turned off if 
atom 2 remains in a superposition of $|0\rangle$ and $|D\rangle$ without any phase accumulation on $|D\rangle$ or population loss into $|B\rangle$. This will be the case provided the probe field is turned on and off adiabatically as compared with $|B\rangle-|D\rangle$ energy splitting, which is equal to the Stark shift $\Delta_S = \tilde \Omega^2/\Delta$ of $|B\rangle$. In the limit $\Omega_c \gg \Omega$, which we will assume from now on, the non-adiabatic coupling between $|D\rangle$ and $|B\rangle$ has an effective Rabi frequency $\Omega_{NA} \sim \Omega/(T \Omega_c)$ \cite{fleischhauer96} giving population loss from the dark state into the bright state of order $\rho_B \sim (\Omega_{NA}/\Delta_S)^2 \sim (\Omega/\Omega_c)^6$ and hence an error of the same order. The errors due to the Stark shift $\Omega^2_{NA}/\Delta_S$ of $|D\rangle$ and due to spontaneous emission are smaller than $(\Omega/\Omega_c)^6$ and $\gamma/\Delta$, respectively. 

In the simplest case, these are the dominant sources of error, so that the total error is
\begin{equation} \label{Pe}
P_e \sim (\gamma/\Delta) +  (\Omega/\Omega_c)^6.
\end{equation}
Plugging in $\Omega^2 \sim \Delta/\tau$ and minimizing with respect to $\Delta$ gives $\Delta \sim  (\gamma \tau^3 \Omega_c^6)^{1/4}$ and $P_e \sim \left[\gamma/(\tau \Omega_c^2)\right]^{3/4}$, which can be made arbitrarily small by increasing control intensity.

However, other sources of error exist. For $d \ll \lambda$, dipole-dipole interactions and cooperative decay effects may become important \cite{guo95}. Cooperative decay will not qualitatively change the errors since the desired evolution is close to unitary. Assuming that we have only two atoms and that $d \ll \lambda$, taking the axis of quantization to coincide with the $x$-axis, the dipole-dipole Hamiltonian can be written as $H_{dd}=(\vec \mu_1 \cdot \vec \mu_2 - 3 (\vec \mu_1 \cdot \hat x) (\vec \mu_2 \cdot \hat x))/d^3$, where $\vec \mu_i$ is the electric dipole operator of the $i$th atom. Since most of the population will stay in $|0\rangle$ and $|1\rangle$, the dipole-dipole interactions involving state $|r\rangle$ can be ignored. Then, provided $|0\rangle-|e\rangle$ and $|1\rangle-|e\rangle$ have different polarizations or sufficient frequency difference, $H_{dd} \approx -g_{0} (|0e\rangle \langle e0| + |e0\rangle \langle 0e|) - g_{1} (|1e\rangle \langle e1| + |e1\rangle \langle 1e|)$, where $|\alpha \beta\rangle$ denotes a two-atom state with atom 1 in $|\alpha\rangle$ and atom 2 in $|\beta\rangle$ and where $g_{0}$ and $g_{1}$ are proportional to $g = \gamma/(k d)^3$ with proportionality constants that depend on the polarizations of the transitions. Then a perturbative calculation shows that dipole-dipole interactions introduce an error $\sim (\Omega g/(\Omega_c \Delta))^4$ \cite{dipolenote}.  

Additional errors are associated with imperfections in the control field node and with finite localization of atoms. If atom 1 was perfectly localized at a single point, a residual control field at the node ($\Omega_c(0) \neq 0$) would result in population $(\Omega_c(0)/\Omega)^2$ in the dark state $|D\rangle$ (now defined for atom 1). However, even if $\Omega_c(0) = 0$, atom 1 can still interact with the control field due to finite extent $a_0$ of its wave-function. Assuming $\Omega_c(0) \lesssim \Omega_0 k' a_0$ \cite{nodenote}, the error due to finite atomic extent (discussed below) will dominate over $(\Omega_c(0)/\Omega)^2$. 

To analyze the problem of localization for atoms in optical lattices and ions in linear Paul traps, we assume that atom 1 sits in the ground state of a harmonic oscillator potential with frequency $\omega$ and, therefore, has spread $a_0 = \sqrt{\hbar/(2 m \omega)}$, where $m$ is the mass of the atom, as shown schematically in Fig.\ \ref{rotations}(c). We assume $\Omega_c(x) = \Omega_0 k' x = \Omega_{ca} (\hat a^\dagger + \hat a)$, where $\Omega_{ca} = \Omega_c(a_0)$ and $\hat a$ is the oscillator annihilation operator. $\Omega_c(x)$ will then couple $|e,n\rangle$ and $|r,m\rangle$ only when $n = m \pm 1$, where $|\alpha, n\rangle$ denotes atom 1 in internal state $|\alpha\rangle$ in $n$th harmonic level. The dominant error can be estimated by keeping only states $|1, 0\rangle$, $|e,0\rangle$, and $|r, 1\rangle$. A perturbative calculation shows that the two limits, in which the error is small are: (a) fast limit $\omega \tau \lesssim 1$, in which case $P_e \sim (\Omega_{ca}/\Omega)^2$; (b) adiabatic limit $\omega \tau \gg 1, (\Omega_{ca}/\Omega)^2$, in which case a small change in the Stark shift of $|1, 0\rangle$ can be compensated by slightly adjusting $\tau$ to yield $P_e \sim (\Omega_{ca}/\Omega)^2/(\tau \omega)^4$.

For atom 2 centered at $x = d$, we have $\Omega_c(x) = \Omega_0 k' d + \Omega_c k' (x-d)$, i.e.\ the desired coupling $\Omega_c$ within each harmonic level is accompanied by coupling of strength $\sim \Omega_{ca}$ between different harmonic levels. Numerical simulations show that provided $\Omega_{ca} < 0.1\, \Omega_c$ (which will always hold below), this coupling has an insignificant effect. 

The error budget for the single-qubit phase gate is summarized in Table \ref{errors}. In general, for a given set of experimental parameters, using $\Omega^2 \sim \Delta/\tau$ to eliminate $\Omega$ in favor of $\Delta$, one has to write the total error as the sum of the errors in Table \ref{errors} and minimize it with respect to $\Omega_0$ and $\Delta$ (we assume $\Omega_0/2 \pi \leq 1$ GHz). We will illustrate this procedure for three systems: ions, solid-state qubits, and neutral atoms. Since ion and neutral atom examples will have $d \sim \lambda$, we take $\Omega_c = \Omega_0$ for them, while for solid-state qubits, we take $\Omega_c = \Omega_0 k' d$. We take $\Omega_{ca} = \Omega_0 k' a_0$, except for neutral atoms, as discussed below. We note that stimulated Raman transitions \cite{wineland98}, resulting in qubit rotations, can also be treated in exactly the same way, yielding similar error probabilities. Moreover, this error analysis is readily extendable to spatially selective qubit measurements and optical pumping, as well as to dipole-dipole two-qubit gates for qubits separated by $d \ll \lambda$.

\begin{table}
\begin{tabular}{|c|l|l|}
\hline
&Error source & Error scaling ($P_e$) \\
\hline \hline
1&decay error on atom 1 & $\gamma/\Delta$ \\
\hline
&\textit{localization error on atom 1:} & \\
\hline
2& - ions and atoms in fast limit &  $(\Omega_{ca}/\Omega)^2$ \\
& and solid-state qubits \cite{solidnote} & \\
\hline
3& - ions and atoms in adiabatic limit & $(\Omega_{ca}/\Omega)^2/(\tau \omega)^4$ \\
\hline
4& unitary error on atom 2 &  $\left(\Omega/\Omega_c \right)^6$\\
\hline
5& dipole-dipole error & $\left(g \Omega/(\Delta \Omega_c)\right)^{4}$ \\
\hline
6& $|r\rangle$ decay on atom 2 for Rb & $(\Omega/\Omega_c)^2 \gamma_r \tau$ \\
\hline
\end{tabular}
\caption{Error budget for the single-qubit phase gate.} \label{errors}
\end{table}

Several approaches to control field node creation exist. One or two standing waves can be used to generate planes or lines, respectively, of zero field with field amplitudes scaling linearly near the zeros. If one has a regular array of atoms (e.g.\ in an optical lattice), arrays of zeros can be chosen to have spacing incommensurate or commensurate with atomic spacing, allowing to address single or multiple atoms, respectively. One can also create control field nodes using holographic techniques \cite{grier02}, which allow one to generate single optical vortices (such as in a Laguerre-Gaussian beam) or an arbitrary diffraction-limited two-dimensional array of them. For simplicity, we consider the case when atoms are sensitive only to one polarization of the control field (e.g.\ if a magnetic field is applied to remove degeneracies). Then the quality of a standing wave node in this polarization component is determined by the interference contrast, which is limited by the mismatch between the amplitudes of this component in the two interfering waves. On the other hand, in an optical vortex, if the phase of the desired polarization component picks up a nonzero multiple of $2 \pi$ around a closed loop, for topological reasons this loop must enclose a line (in three dimensions) where the amplitude of this polarization component exactly vanishes (see e.g.\ \cite{nye99dennis03}). Furthermore, the Rabi frequency in an optical vortex rises radially from the center as $|\Omega_c(x)| \sim \Omega_0 (x/w)^l$, where $w \gtrsim \lambda'$ is the beam waist and the topological charge $l$ is a positive integer. Therefore, in some cases, the use of vortices with $l > 1$ instead of standing waves or $l = 1$ vortices can improve the resolution by decreasing the undesired coupling of the control to atom 1. We will use an $l = 2$ vortex for the neutral-atom example, in which case we take $\Omega_{ca} = \Omega_0 (k' a_0)^2$ in error $\#2$ in Table \ref{errors}.

We first analyze ions in linear Paul traps. We consider for concreteness $^{40}$Ca$^+$ \cite{mcdonnell04} with  $|0\rangle = |4 S_{1/2}, m\! =\! 1/2\rangle$, $|1\rangle = |4 S_{1/2}, m\! =\! -1/2\rangle$, $|e\rangle = |4 P_{1/2}, m \!=\! 1/2\rangle$, and $|r\rangle = |3 D_{3/2}, m \!=\! 3/2\rangle$. Note that $\lambda = 397$ nm and $\lambda' = 866$ nm are far enough apart to ignore off-resonant cross coupling. Then, for $\tau = 1$ $\mu$s, $\omega/2 \pi = 10$ MHz, and $d = 1-3$ $\mu$m, errors $\#1$ and $\#4$ from Table \ref{errors} form the dominant balance, so that Eq.\ (\ref{Pe}) applies and $P_e \sim \left[\gamma/(\tau \Omega_c^2)\right]^{3/4}$, which is $\sim 10^{-4}$ for $\Omega_0/2 \pi = 1$ GHz (with optimal $\Delta/(2 \pi) \sim 200$ GHz and $\Omega/(2 \pi) \sim 200$ MHz). This and the next two error estimates are significantly lower than the errors produced by naive probe focusing.    

For solid-state qubits (e.g.\ Nitrogen-Vacancy color centers in diamond \cite{childress06}), we take $a_0 = 0.5$ nm, $\lambda = \lambda' = 700$ nm, $\gamma/2 \pi = 5$ MHz, and $\tau = 1$ $\mu$s, which, for $d$ between $100$ nm and $20$ nm, makes errors $\#2$ and $\#4$ form the dominant balance, so that $P_e \sim (a_0/d)^{3/2}$ is between $5 \times 10^{-4}$ and $5 \times 10^{-3}$. For $d < 10$ nm, $\Omega_0/2 \pi = 1$ GHz is insufficient to suppress the dipole-dipole error (error $\#5$ in Table \ref{errors}), and the gate fidelity sharply drops.

To analyze atoms in optical lattices, we consider  
$^{87}$Rb with $|0\rangle = |5 S_{1/2}, F \!=\! 2, m_F \!=\! 2\rangle$, $|1\rangle = |5 S_{1/2}, F \!=\! 1, m_F \!=\! 1\rangle$, $|e\rangle = |5 P_{1/2}, F \!=\! 2, m_F \!=\! 2\rangle$, and $|r\rangle = |4 D\rangle$.  
$|4 D\rangle$ decays with rate $2 \gamma_r = 1/(90\textrm{ ns})$; so to reduce the error $\sim \rho_r \gamma_r \tau \sim (\Omega/\Omega_c)^2 \gamma_r \tau$ on atom 2 (error $\#6$ in Table \ref{errors}), we choose short $\tau = 10$ ns. For $\omega/2 \pi = 50$ kHz and $\Omega_0/2 \pi = 1$ GHz, errors $\#2$ and $\#6$ form the dominant balance, so that $P_e \sim (\Omega_{ca}/\Omega_{c}) \left(\tau \gamma_r\right)^{1/2} \sim 0.01$. This error can be further reduced by tightening the traps for the duration of the gate either by increasing the power of or by decreasing the detuning of the lattice beams. 

Our selective addressability technique has several advantages that may enable it to outperform alternative all-optical addressability proposals based on the gradient method \cite{thomas93zhang06lee07}. First, the nonlinear response provided by the dark states may potentially provide our method with superior error scaling. Second, in the gradient method, the control field typically couples states that are populated at some point during 
the gate. In contrast, in our method, the control field is small (ideally, vanishing) on the atom that is being manipulated, while on the neighboring atoms the population of level $|r\rangle$ (coupled by the control to level $|e\rangle$) is negligible and becomes smaller as the control power grows. As a result, in contrast to the gradient method, our method (1) avoids unwanted forces on atoms due to Stark shift gradients [and hence prevents unwanted entanglement of external and internal degrees of freedom] and (2) avoids excessive spontaneous emission, which may take place if the control field mixes populated stable states with short-lived excited states.

We now outline some new avenues opened by the coherent selective addressability technique. Although we discussed in detail only the application of this technique to selective phase gates (equivalently, Raman transitions), it has obvious generalizations to geometric gates \cite{duan01}, fluorescence detection, and optical pumping/shelving, as well as to the generation (in combination with dipole-dipole interactions and assuming $d \ll \lambda$) of entangling gates between atoms. In addition to the applications to atoms in optical lattices, to ions in linear Paul traps, and to solid-state qubits, our technique may also allow for single-atom addressability in recently proposed sub-wavelength optical lattices \cite{daley07}. Moreover, a combination of similar ideas involving dark states and the nonlinear atomic response can itself be used for creating deep sub-wavelength-separated traps and flat-bottom traps. Finally, better optimization (e.g.\ using optimal control theory to shape laser pulses) can further reduce the errors. Therefore, we expect this technique to be of great value for fields ranging from  quantum computation and quantum simulation to coherent control, all of which can benefit from high-fidelity addressability at $d \lesssim \lambda$.

We thank  D.E.\ Chang, A.\ Peng, J.\ Gillen,  T.\ Calarco, S. F\"olling, J.E.\  Thomas, and M.R.\ Dennis for 
discussions. This work was supported by the 
NSF, Harvard-MIT CUA, Packard Foundation, and AFOSR MURI. 
P.Z.\ acknowledges support by the Austrian Science Foundation and the EU.

Note added: after completing this work, we became aware of  related proposals \cite{cho07, yavuz07, juzeliunas07} to use dark state position dependence to achieve sub-wavelength resolution.


\begin{thebibliography}{99}
\bibitem{dipolenote} If $\Omega_c$ is not sufficiently larger than $\Omega$, $P_e \sim (\Omega g/(\Omega_c \Delta))^2$. However, $(\Omega g/(\Omega_c \Delta))^2$ and $(\Omega g/(\Omega_c \Delta))^4$ depend on $d$ so sharply that they give almost the same resolution.
\bibitem{nodenote} The condition is modified in the adiabatic limit (see error $\#3$ in Table \ref{errors}): $\Omega_c(0) \lesssim \Omega_0 k' a_0 /(\omega \tau)^2$.
\bibitem{solidnote} For solid state qubits, we model imperfect localization by replacing $\Omega_c(0)$ with $\Omega_{ca} = \Omega_c(a_0)$ to give an error $\sim (\Omega_{ca}/\Omega)^2$, where $a_0$ is the electronic-wavefunction extent.
\end{thebibliography}
\end{document}